\documentclass[amsfonts,floatfix,aps,pre,twocolumn,10pt]{revtex4}
\usepackage{array}
\usepackage{amsmath}
\usepackage{graphicx}
\newcommand{\ind}[1]{\textrm{\tiny{#1}}}
\renewcommand{\phi}{\varphi}
\renewcommand{\d}{\textrm{d}}
\newcommand{\eww}[1]{\langle#1\rangle}
\newcommand{\phisol}{\phi_\ind{sol}}

\newcommand{\psol}{p_\ind{sol}}

\newcommand{\pact}{p_\ind{act}}

\newcommand{\taustar}{\tau^*}
\newcommand{\Eeff}{E_\ind{eff}}
\newcommand{\Dest}{D_\ind{est}}

\newcommand{\epsmb}{\epsilon}
\newcommand{\kB}{k_\ind{B}}
\newcommand{\Tc}{T_\ind{c}}
\newcommand{\Tg}{T_\ind{g}}

\newcommand{\BMNIP}{Ni$_{80}$P$_{20}$}
\newcommand{\sigmaref}{\sigma_\ind{ref}}
\newcommand{\Tref}{T_\ind{ref}}
\newcommand{\tref}{t_\ind{ref}}
\newcommand{\Dref}{D_\ind{ref}}
\newcommand{\epsmin}{\epsilon_\ind{min}}
\newcommand{\epsref}{\epsilon_\ind{ref}}
\newcommand{\equref}[1]{Eq.\ref{#1}}

\newcommand{\FIG}[1]{Fig.\ref{#1}}
\newcommand{\wideth}{7.5cm}

\begin{document}
\title{What does the potential energy landscape tell us about the
dynamics of supercooled liquids and glasses?}
\author{B. Doliwa$^1$ and
A. Heuer$^2$}
\affiliation{$^1$Max Planck Institute for Polymer Research, Mainz, Germany}
\affiliation{$^2$Institute of Physical Chemistry, University of M\"unster -
  M\"unster, Germany}
\date{\today}
\begin{abstract}
For a model glass-former we demonstrate via computer simulations how
macroscopic {\it dynamic} quantities can be inferred from a PEL
analysis. The essential step is to consider whole superstructures of
many PEL minima, called metabasins, rather than single minima.  We
show that two types of metabasins exist: some allowing for quasi-free
motion on the PEL (liquid-like), the others acting as traps
(solid-like).  The activated, multi-step escapes from the latter
metabasins are found to dictate the slowing down of dynamics upon
cooling over a much broader temperature range than is currently
assumed.
\end{abstract}
\maketitle

The theoretical understanding of supercooled liquids and glasses
is one of the important unsolved problems of statistical physics.
Since the pioneering work of Goldstein~\cite{Goldstein:1969}, it
has been realized that the potential energy landscape (PEL)
viewpoint is useful for characterizing supercooled liquids and
glasses.  To this end, one considers the high-dimensional vector
of all particle coordinates as a point moving on the surface of
the total potential
energy~\cite{Stillinger:333,Debenedetti:217,Sastry:209}.  At
sufficiently low temperatures, the system resides near the local
minima of the high-dimensional PEL.  It has turned out that the
PEL description of thermodynamics starts to work when cooling
below approximately $T=2\Tc$, where $\Tc$ is the critical
temperature of mode-coupling theory~\cite{Gotze:1992}. In this
temperature regime the statistical properties of PEL minima fully
determine all {\it thermodynamic}
properties~\cite{Sastry:198,Sciortino:59,Sciortino:280,Buchner:11,Emilia:302}.

What is still lacking in the literature is a full quantitative
understanding of {\it dynamics}, i.e. of the slowing down of
molecular motion upon cooling.  In this connection, the
single-particle diffusion coefficient, $D(T)$, is a typical
macroscopic quantity of interest.  In general, knowledge of
thermodynamic properties is not sufficient to predict $D(T)$.
However, the validity of the Adam-Gibbs relation~\cite{Adam:39},
connecting $D(T)$ to the configurational entropy, suggests the
presence of a strong link between thermodynamics and
dynamics~\cite{Saika:247,Scala:71,Starr:140}. For a model
glass-former we have demonstrated via computer simulations that
$D(T)$ can indeed be inferred from a PEL analysis in a
quantitative way. The essential step, however, is to consider
whole superstructures of many PEL minima, called
metabasins~\cite{Stillinger:333,Buchner:11,doliwa:392,Reichman:398,doliwa:341,Middleton:214},
rather than single minima.  Metabasins are reminiscent of
protein-folding funnels~\cite{Bryngelson} or related structures
in small clusters~\cite{Ball:379}.

The goal of this paper is to derive physical consequences of this
mapping between PEL and dynamics and thus to obtain a coherent
picture of the glass transition. In particular we will dwell on
the interpretation of the activation energies for fragile systems
in the Arrhenius diagram, the concept of liquid-like and
solid-like behavior~\cite{Gartenzwerg}, and the question of a
possible crossover at $T_c$, as implied by the mode-coupling
temperature and the analogy to $p$-spin models.

In what follows we analyze a binary mixture of Lennard-Jones
particles (BMLJ). All details of the BMLJ have been
reported in~\cite{doliwa:392}. This system is designed to model a
\BMNIP~mixture.  Langevin molecular dynamics have been employed
with a simulation box of $N=65$ particles at density
$\rho=1.2\sigmaref^{-3}$. Throughout the paper, all quantities are
given in reduced units, which, in the case of \BMNIP, correspond
to $\sigmaref=2.218$~\AA, $\epsref=7.762$~kJ/mol,
$\tref=1.323$~fs, $\Tref=934$~K, and
$\Dref=\sigmaref^2/\tref=0.372~\textrm{cm}^2/s$.  For the
temperature regime of our analysis, a system size of $N=65$ is
provably sufficient to reproduce the dynamic properties of a
macroscopic system~\cite{Buchner:193,Doliwa:404}: We have
checked, e.g.,  that $D(T)$ is the same for $N=130$ within 10\%
for $T>\Tc=0.45\Tref$. The minima are obtained by a steepest descent
minimization. As usual we perform regular quenches of the system
to monitor the energies of the corresponding energy minima.
In addition, we use temporal interval bisectioning to resolve
the elementary transitions between minima~\cite{doliwa:341}.

The definition of metabasins (MBs) is motivated by the observation
that during the molecular dynamics in configuration space the system
performs several back-and-forth jumps between adjacent minima until
finally this region is left. The description of the transport may be
simplified if these minima are regarded as a single superstructure,
i.e. a MB~\cite{doliwa:341}. Thus, the MBs correspond to an
appropriate tiling of configuration space.  Although irrelevant
for the understanding of thermodynamic properties, they are of
outstanding importance for dynamics. As shown in~\cite{Saksaengwijit}
it is possible to find a strict definition of MBs which can be also
used in practice.  Similarly to the previous consideration of energy
minima the time evolution of the system may be regarded as a
continuous sequence of MB visits with individual residence (waiting)
times~$\tau$.


We briefly summarize the main results of our analysis which
will be important throughout the paper. They all hold
for $T<2T_c$. (i)~Analyzing the escape characteristics from MBs
by repeated simulations from the same MBs at different
temperatures~$T$~\cite{doliwa:392}, it turns out that the mean
residence time in MBs of energy $\epsmb$ is given to a good approximation by
\begin{equation}
\label{eqtau}
\langle\tau(\epsmb,T)\rangle\approx\tau_0(\epsmb)\exp(E(\epsmb)/\kB T).
\end{equation}
The Arrhenius form of $\langle\tau(\epsmb,T)\rangle$ suggests that
the escapes from MBs are thermally activated, with a barrier
height of $E(\epsmb)$, introduced here as a fitting parameter. The
prefactor $\tau_0(\epsmb)$ turns out to be basically independent
of $\epsmb$. (We define $\epsmb$ as the energy of the lowest
minimum within a MB.) (ii)~$E(\epsmb)$ is directly related to the
PEL barriers in the high-dimensional surrounding of the MBs. By a
detailed discussion of this issue, incorporating the funnel-like
nature of MBs, this was verified~\cite{doliwa:392}. Thus, by
analyzing the local topology of MBs, $E(\epsmb)$ can be predicted.
We may interpret $E(\epsmb)$ as the depth of a MB of energy
$\epsmb$. The deeper the MB in the PEL, i.e. the lower $\epsmb$,
the higher the activation energy. In this analysis one has to take
into account that the escape from a MB is a multi-step process,
i.e. generally comprises hops between several minima.  (iii)~A
crucial quantity is the average residence time
$\langle\tau(T)\rangle$. It is defined as the average over all MBs
encountered at a specific temperature, i.e.
$\eww{\tau(T)}=\int\d\tau\tau\phi(\tau,T)$.  By $\phi(\tau,T)$ and
$\phi(\epsmb,T)$, respectively, we denote the distribution of
waiting times and energies of visited MBs. (iv)~The {\it
population} of MBs with energy~$\epsmb$ is given by
$p(\epsmb,T)=\eww{\tau(\epsmb,T)}\phi(\epsmb,T)/\eww{\tau(T)}$.
Numerically, it is indistinguishable from the population of
minima, normally studied in this field and is purely gaussian in
the accessible energy range~\cite{doliwaPhD}. Using \equref{eqtau}
the inverse average waiting time can be written as
\begin{eqnarray}
\label{eqtau2} \tau_0/\langle \tau(T) \rangle\approx\int\d\epsmb
p(\epsmb,T) \exp(-E(\epsmb)/\kB T)\nonumber\\
\equiv \exp(-\Eeff(T)/\kB T),
\end{eqnarray}
where the energy dependence of $\tau_0(\epsmb)$ has been neglected.
$\Eeff(T)$ can be interpreted as the typical barrier height encountered at
temperature $T$. (v)~A simple relation exists between $\eww{\tau(T)}$
and $D(T)$ via
\begin{equation}
\label{eqd} D(T)\approx\frac{a^2}{6N \langle\tau(T)\rangle},
\end{equation}
with a temperature-independent effective jump length
$a\approx1.0\sigmaref$. Thus, the temperature dependence of $D(T)$
is {\it exclusively} determined by the average waiting time,
whereas all spatial aspects of diffusion are temperature
independent. Such a simple relation does not hold on the level of
single minima. Actually, as a side effect it turns out that the
dynamics can be basically described as a random-walk between MBs,
thus suggesting a non-topographic view. With the definition $D_0
\equiv a^2/6N\tau_0$ we can thus write
\begin{equation}
\label{eqfinal} D(T)\approx D_0\exp(-\Eeff(T)/\kB T).
\end{equation}
Note that the ingredients of the constant $D_0$, i.e. $a$ and
$\tau_0$, have been obtained from the simulations mentioned above.
In the units of our simulation we have $D_0\equiv
a^2/6N\tau_0\approx1.3\cdot 10^{-5}\Dref$.  If the BMLJ is mapped
on a \BMNIP~alloy this corresponds to
$D_0\approx4.8\times10^{-6}~\textrm{cm}^2/\textrm{s}$.

\begin{figure}[!ht]
\includegraphics[width=\wideth]{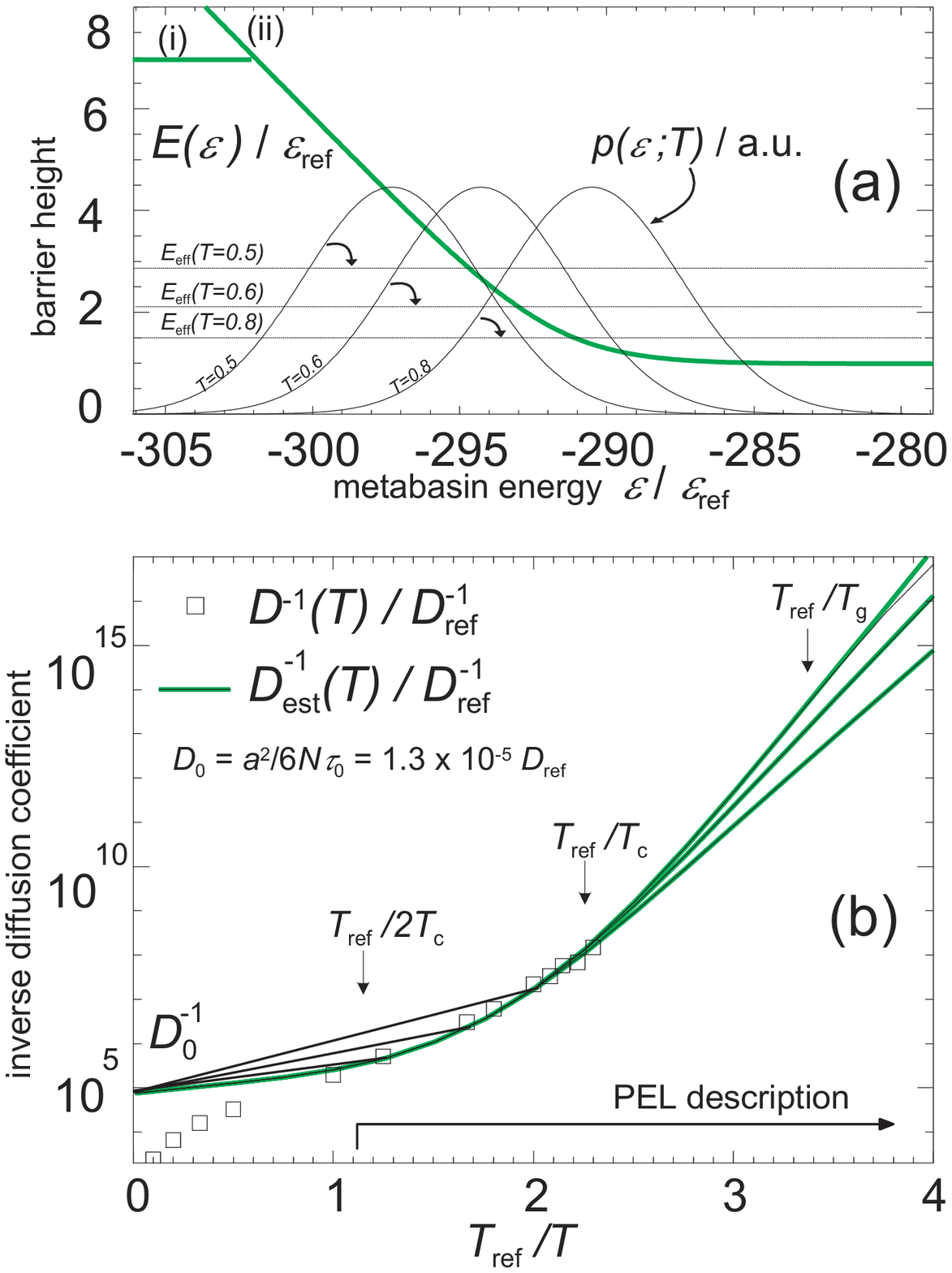}
\caption{ \label{prl1} Estimating the diffusion coefficient from
the depths of MBs and their distribution.  (a) The barrier height
$E(\epsmb)$ as a function of MB energy, and the population of MB
energies, $p(\epsmb,T)$, for $\epsmin=-306\epsref$ at different
temperatures. $p(\epsmb,T)$ is purely gaussian~\cite{Buchner:191}.
(b) Comparison of $\Dest(T)$ with  $D(T)$. The three curves
correspond to different lower PEL cutoffs $\epsmin/\epsref=-302,
-304$, and $-306$, from bottom to top. Different extrapolations of
$E(\epsmb)$ to $\epsmb<-302\epsref$ (see (a)) lead to nearly
identical $\Dest(T)$'s (no visible difference between the curves
corresponding to (i) and (ii) in (a)). We set
$\Tg\approx0.3\Tref$~\cite{Sciortino:59}.  }
\end{figure}

Here we discuss the relevant physical implications of
Eqs.\ref{eqtau}-\ref{eqfinal}. In \FIG{prl1}(a) we show the
barrier height $E(\epsmb)$ as a function of MB energy, and the
population of MB energies, $p(\epsmb,T)$, for different
temperatures. From these quantities we obtain an estimate for the
diffusion coefficient, denoted $\Dest(T)$. A comparison of the
estimated with the true diffusion coefficient $D(T)$ (directly
from molecular dynamics) is shown in~(b). The very good agreement
at $\Tc\le T\le 2\Tc$ reflects the consistency of our approach and
shows that we can indeed understand the macroscopic dynamics from
knowledge of the thermodynamics and the local barriers.

 From the simulations, we know $p(\epsmb,T)$ and $E(\epsmb)$
for $\epsmb>-302\epsref$. Both quantities do not display
finite-size effects in this energy range \cite{Doliwa:404}. Due to
the Boltzmann weighting in Eq.\ref{eqtau2} one may hope that the
available information is sufficient to reliably estimate
$\Dest(T)$ also for $T < T_c$. To check this hypothesis, we have
calculated $\Dest(T)$ thereby using extreme candidates for
possible $\epsmb$-extrapolations ($\epsmb<-302\epsref$); see
\FIG{prl1}. It turns out that $\Dest(T)$ is very insensitive to
$E(\epsmb)$.  For $p(\epsmb;T)$ we have extended the gaussian
description but have set $p(\epsmb;T)=0$ below different
cutoff-energies $\epsmin$. (A lower bound for $\epsmin$ is given by
the condition that the number density of minima is of the order of
$1/\epsref$ ~\cite{Sastry:198}.  This leads to
$-306\epsref\le\epsmin\le-302\epsref$~\cite{doliwaPhD}.) As shown in
\FIG{prl1} (b), the resulting uncertainty in the estimation of
$\Dest(T)$ is small up to $T>0.35\Tref$ and becomes significant for
lower temperatures. Since a cutoff at $-302\epsref$ is probably far
too drastic (see \cite{doliwaPhD} for strong indications), however,
the slowing down can be thus predicted semi-quantitatively down to $T
\approx T_g$, a region which very likely will never be accessible by
molecular dynamics simulations.

Moreover, \equref{eqfinal} implies that the activation energy which is
obtained from connecting $(1/T=0/\Tref,D=D_0)$ and $(1/T,D(T))$ in
an Arrhenius plot by a straight line has a simple interpretation:
it is the typical barrier height the system experiences at a given
temperature~\cite{DyreX}.  This non-trivial interpretation has
very recently been confirmed by hyperquench
experiments~\cite{Angell:ESpray}, where the typical barrier height
was measured by probing the specific heat during the reheating
process.

\begin{figure}[!ht]
\includegraphics[width=\wideth]{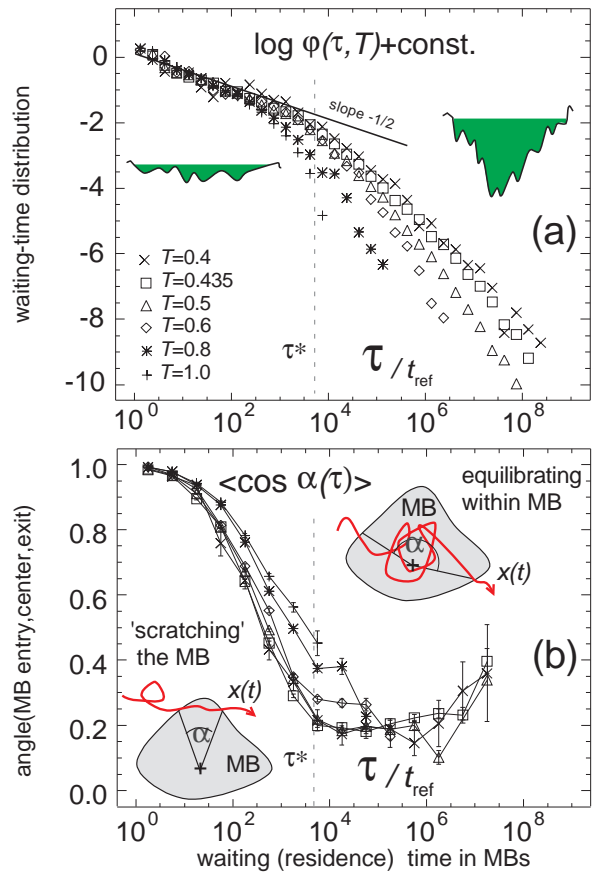}
\caption{
\label{prl2}
Different exploration of MBs, as evidenced by the waiting-time distributions
and the correlations between MB-entry and exit points.
(a)~The distribution $\phi(\tau,T)$ of waiting times at different
temperatures.
(b)~Average value of $\langle\cos\alpha(\tau)\rangle$ over MBs of lifetime
$\tau$ (see text and the sketches in (b), $x(t)$ symbolizes the high-dimensional
trajectory of the system).
}
\end{figure}
To obtain a more detailed picture of MBs we have determined the
distribution of waiting times, $\phi(\tau,T)$, as shown in
\FIG{prl2}(a).  By visual inspection, one can distinguish two
different time regimes (separated by the vertical line at
$\taustar\approx5000\tref$).  In terms of energetics one expects
that the long $\tau$'s arise from deep traps where the system is
caught for a long time. In the opposite limit one may imagine that
there are quite shallow MBs which do not strongly confine the
system so that it will mainly stay close to the high-dimensional
boundary of these MBs (see the sketches in (a)). Thus, short
waiting times correspond to just scratching the MBs. This
expectation can be verified by analyzing the trajectory of the
system during the MB visits; an example of such a computation is
given in \FIG{prl1}(b). There we plot
$\langle\cos\alpha(\tau)\rangle$ over MBs of lifetime $\tau$,
where $\alpha$ is the angle between the entry point, the lowest
minimum of the MB and the exit point of the trajectory (see the
sketches in (b)).  Again, two time regimes with a
temperature-independent crossover time $\taustar$ can be
identified. Short visits to MBs lead to small values of $\alpha$,
meaning that the system indeed merely scratches these MBs.  For
$\tau>\taustar$ the value of $\langle\cos\alpha(\tau)\rangle$
reaches a limiting value of ca.~0.2 , indicating that entry and
exit points are largely uncorrelated.  This should be the case
after a long equilibration inside a MB with many possible exits.
Due
to the
difference in stability, we call the MBs with $\tau>\taustar$ {\it
solid-like}, the other MBs {\it liquid-like}. This notation has been
borrowed from two-state models where these two types of configurations
have been postulated~\cite{Gartenzwerg}.

\begin{figure}[!t]
\includegraphics[width=\wideth]{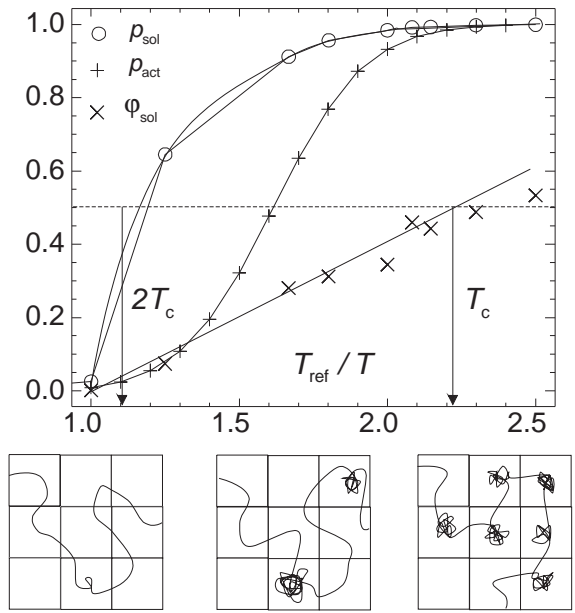}
\caption{
\label{prl3}
The temperature dependence of $\psol$ (the fraction of time spent in
solid-like configurations), $\phisol$ (the fraction of solid-like
configurations encountered during the time evolution), and $\pact$ (a
measure for the contribution of activated processes to the diffusion
coefficient; see text). In the lower part we depict schematic plots of
the scenarios in the three temperature regimes. The squares symbolize
the different MBs.  }
\end{figure}

Clearly, the molecular slowing down upon cooling is caused by the
enhancement of solid-like configurations. This can be quantified in
two different ways. Firstly, one may determine the fraction of
solid-like configurations the system encounters, i.e.
$\phisol(T)\equiv\int_{\taustar}^\infty d\tau\phi(\tau,T)$, where
$\phisol<0.5$ implies that more liquid-like than solid-like
configurations are visited.  Secondly, we can specify the fraction of
time spent in solid-like configurations, which can be expressed as
$\psol(T)\equiv\int_{\taustar}^\infty\d\tau p(\tau,T)$ with $p(\tau,T) =
\tau\phi(\tau,T)/\eww{\tau(T)}$. $\psol>0.5$ implies that the system
is mostly residing in solid-like configurations. In \FIG{prl3} we show
the temperature dependence of $\phisol$ and $\psol$. Three different
temperature regimes can be distinguished (see the sketch in
\FIG{prl3}): For $T>2\Tc$ both quantities are smaller than 0.5. Thus,
the system behaves liquid-like. Interestingly,
this temperature regime (defined by {\it dynamics})
is exactly the
temperature regime for which the minima no longer influence the
{\it thermodynamic} properties of the system~\cite{Sastry:209}.
Below $2\Tc$, $\psol$ is
larger than~0.5.  Thus, the system mainly resides in solid-like MBs.
Finally, below a temperature near $\Tc$, also $\phisol$ exceeds 0.5,
i.e. we have a trap-to-trap motion. This crossover, however, is very
gradual.

With \equref{eqtau2} we can also analyze the question in which
temperature regime the dynamics are dominated by activated
processes. We call a process activated if the activation energy
$E(\epsmb)$ is larger than $5\kB T$. To this end, we consider the
fraction of $\eww{\tau(T)}$ which is made up by activated hops, i.e.
$$\eww{\tau(T)}^{-1}\int_{E(\epsmb)>5\kB T}\d\epsmb\eww{\tau(\epsmb,T)}\phi(\epsmb,T)\equiv\pact(T),$$
as shown in \FIG{prl3}. The crossover temperature for which $\pact =
0.5$ is close to $1.5\Tc$. Thus, already significantly above $\Tc$,
the dynamics are dominated by activated processes.  Please note that
the MBs contributing to the above integral are solid-like since they
fulfill $\eww{\tau(\epsmb,T)}\gtrsim\taustar$ (see~\cite{doliwa:392}).
This is why $\pact(T)\le\psol(T)$.


The present results seem to be at variance with the current
understanding that activated transitions between PEL minima set in
at $\Tc$, whereas above $\Tc$ they are irrelevant.  The latter
scenario is motivated by the properties of $p$-spin
models~\cite{Crisanti:368} and has been backed by the observations
that the number of free directions (obtained via
instantaneous-normal-mode
analyses~\cite{LaNave:265,Chowdhary:334}), is directly related to
the diffusion coefficient and that above $\Tc$ the system is close
to transition states rather than close to
minima~\cite{Angelani:215,Broderix:228,Grigera:264}.  In our
earlier publication~\cite{doliwa:392}, we have already pointed out
the technical shortcomings of the latter kind of investigations.
More importantly, though, the above-cited works do not incorporate
the MB structure of configuration space.  In doing so, one has
focused on the vast majority of {\it intra}-MB transitions, which,
however, are irrelevant for relaxation.

Finally we note that the results of our work with respect to the
crossover to activated behavior as well as the non-topographic nature
of inter-MB dynamics is compatible with the implications
obtained from a recent analysis of spin-facilitated models of the
glass transition~\cite{Berthier:445}.

We thank C.A. Angell, L. Berthier, J.P. Garrahan, and H.W. Spiess
for helpful discussions.

\bibliographystyle{apsrev}
\bibliography{nat}

\end{document}